\documentclass[preprint,aps,prd,superscriptaddress,nofootinbib]{revtex4}%
\usepackage{graphicx}
\usepackage{amsmath}
\usepackage{amssymb}
\usepackage{bm}
\usepackage{epsfig}
\usepackage{mathrsfs}
\usepackage{hyperref}

\usepackage{soul}
\begin{document}
\preprint{}
\title{\mbox{}\\[10pt]
Addendum: New approach to the resummation of logarithms in
Higgs-boson decays to a vector quarkonium plus a photon\\             
[0pt][Phys.\ Rev.\ D {\bf 95}, 054018 (2017)]
}
\author{Geoffrey~T.~Bodwin}
\email[]{gtb@anl.gov}
\affiliation{High Energy Physics Division, Argonne National Laboratory,
Argonne, Illinois 60439, USA}
\author{Hee~Sok~Chung}
\email[]{hee.sok.chung@cern.ch}
\affiliation{Theory Department, CERN, 1211 Geneva 23, Switzerland}
\author{June-Haak~Ee}
\email[]{chodigi@gmail.com}
\affiliation{Department of Physics, Korea University, Seoul 02841, Korea}
\author{Jungil~Lee}
\email[]{jungil@korea.ac.kr}
\affiliation{Department of Physics, Korea University, Seoul 02841, Korea}
\date{\today}
\begin{abstract}
In this addendum to Phys.\ Rev.\ D {\bf 95}, 054018 (2017)
we recompute the rates for the decays of the Higgs boson to a vector
quarkonium plus a photon, where the vector quarkonium is $J/\psi$,
$\Upsilon(1S)$, $\Upsilon(2S)$, or $\Upsilon(3S)$. We correct an error
in the Abel-Pad\'e summation formula that was used to carry out the
evolution of the quarkonium light-cone distribution amplitude in 
Phys.\ Rev.\ D {\bf 95}, 054018 (2017). 
We also correct an error in the scale of
quarkonium wave function at the origin in 
Phys.\ Rev.\ D {\bf 95}, 054018 (2017) and
introduce several additional refinements in the calculation.
\end{abstract}
\maketitle


In Ref.~\cite{Bodwin:2016edd}, rates for the decays of the Higgs boson
to a vector heavy quarkonium plus a photon ($H\to V+\gamma$) were
computed by making use of the light-cone amplitude for the direct 
decay process. In the direct process, the Higgs boson decays into a
heavy-quark-antiquark pair ($Q\bar Q$), which radiates a gluon and
evolves into a vector heavy quarkonium. In this addendum to 
Ref.~\cite{Bodwin:2016edd}, we make several improvements to the 
calculation of the direct amplitude in Ref.~\cite{Bodwin:2016edd} and 
also correct an error in the scale of the heavy-quarkonium wave function 
at the origin and an error in the application of the Abel-Pad\'e method to the 
evolution of the light-cone distribution amplitude (LCDA).

In Ref.~\cite{Bodwin:2016edd}, $\phi_V(x)$, the heavy-quarkonium
light-cone distribution amplitude at light-cone momentum fraction
$x$, was evolved from the initial scale $\mu_0$, which is of order the
heavy-quark mass, to the final scale $\mu$, which is of order the
Higgs-boson mass $m_H$, by making use of the Abel-Pad\'e method. 

As was explained in Ref.~[1], the Abel summation in the 
Abel-Pad\'e method defines generalized functions 
(distributions) in the nonrelativistic QCD (NRQCD) 
expansion of the LCDA at the initial scale $\mu_0$ as 
limits of sequences of ordinary functions.
In Eq.~(31) of Ref.~\cite{Bodwin:2016edd}, the
Abel summation was erroneously applied to $\phi_V(\mu)$, rather than to
$\phi_V(\mu_0)$. Equation~(31) of Ref.~\cite{Bodwin:2016edd} should be
corrected to the following:
\begin{equation}
\label{limit-eigenvalue-sum}
{\cal M}=
\lim_{z\to 1} \sum_{n,m=0}^\infty T_m(\mu) U_{mn}(\mu,\mu_0) 
z^n \phi_{n}^\perp(\mu_0).
\end{equation}
Here, ${\cal M}$ is the light-cone amplitude, $\phi_{n}^\perp(\mu_0)$ is
the $n$th Gegenbauer moment of $\phi_V^\perp(x,\mu_0)$ [Eq.~(19b) of
Ref.~\cite{Bodwin:2016edd}], $ U_{mn}(\mu,\mu_0)$ is the evolution
matrix for the LCDA [Eq.~(B1) of Ref.~\cite{Bodwin:2016edd}], and $T_m$
is the $m$th Gegenbauer moment of the hard-scattering amplitude
[Eq.~(20b) of Ref.~\cite{Bodwin:2016edd}]. The essential change in
Eq.~(\ref{limit-eigenvalue-sum}) is to replace $z^m$ with $z^n$.

Because the evolution matrix in leading logarithmic (LL) order is
diagonal, this change affects the expression for the light-cone
amplitude only at next-to-leading logarithmic (NLL) order and higher. In
Ref.~\cite{Bodwin:2016edd}, only the LCDA at leading order in $\alpha_s$
and $v$ was evolved at NLL order. (Here, $\alpha_s$ is the QCD running
coupling and $v$ is the velocity of the $Q$ or $\bar Q$ in the
quarkonium rest frame.)  Because the sums in
Eq.~(\ref{limit-eigenvalue-sum}) are not absolutely convergent, it is
not obvious that the correction in Eq.~(\ref{limit-eigenvalue-sum})
would not affect the results.  However, it turns out that the correction
in Eq.~(\ref{limit-eigenvalue-sum}) does not shift the numerical results
for the direct amplitudes significantly from those that were reported in
Ref.~\cite{Bodwin:2016edd}, although it does improve the stability of
the convergence of the Abel-Pad\'e procedure.

The light-cone amplitude for the direct process can be written
\cite{Jia:2008ep}, by virtue of NRQCD factorization
\cite{Bodwin:1994jh},  as a sum of products of  short-distance
coefficients with NRQCD long-distance matrix elements (LDMEs)
\cite{Wang:2013ywc}. We denote the NRQCD LDME of leading order in $v$ by
$\langle {\cal O}_1^V\rangle$. The quantity $\langle {\cal
O}_1^V\rangle$ can be written in terms of the quarkonium wave function
at the origin $\Psi_V(0)$ (Ref.~\cite{Bodwin:1994jh}):
\begin{equation}
\langle {\cal O}_1^V\rangle=\frac{1}{2N_{c}}|\Psi_V(0)|^2,
\label{LDME-phi0}
\end{equation}
where $N_c=C_A=3$ for $\textrm{SU}(3)$ color.

For the computations in Ref.~\cite{Bodwin:2016edd}, which involve
$J/\psi$ and $\Upsilon(nS)$ final states, with $n=1$, $2$, or $3$,
values for $\Psi_V(0)$ were inferred from values of $\langle {\cal
O}_1^V\rangle$ that were given in
Refs.~\cite{Bodwin:2007fz,Chung:2010vz}. In those papers, $\langle {\cal
O}_1^V\rangle$ was computed by making use of the NRQCD expansion of the
electronic decay rate $\Gamma(V\to e^+e^-)$ through orders
$\alpha_s(\mu)$ and $v^2$.\footnote{The calculations in
Refs.~\cite{Bodwin:2007fz,Chung:2010vz} also included a class of
corrections of order $v^4$ and higher. However, there are additional
corrections of order $v^4$ that were not considered in those papers. In
order to streamline the discussion, we show explicitly only
contributions through order $v^2$ in the present paper.}  That
expansion can be obtained from Eq.~(\ref{LDME-phi0}), the relation
\begin{equation}                                                      
\Gamma(V\to e^+e^-)=\frac{4\pi}{3m_V}\alpha^2(m_V)e_Q^2     
f_V^{\parallel 2},
\label{EM-width}
\end{equation}
and the NRQCD expansion
\begin{equation}                                                 
f_V^\parallel                                                 
=
\frac{\sqrt{2N_c}\sqrt{2m_V}}{m_V}\Psi_V(0)
\left[1-\frac{1}{6}\langle v^2\rangle_V
-8\frac{\alpha_s(\mu) C_F}{4\pi}
+O(\alpha_s^2,\alpha_s v^2,v^4)
\right].
\label{fVpar-expn}
\end{equation}
Here, $f_V^\parallel$ is the decay constant for a longitudinally
polarized quarkonium, $\alpha(\mu)$ is the electromagnetic
fine-structure constant at the scale $\mu$, $e_Q$ is the heavy-quark
charge, $m_V$ is the quarkonium mass, $C_F=(N_c^2-1)/(2N_c)$, and
$\langle v^2\rangle_V$ is the ratio of the NRQCD decay LDME of relative
order $v^2$ to the NRQCD decay LDME of leading order in $v$ ($\langle
{\cal O}_1^V\rangle$). (See Eq.~(6) of Ref.~\cite{Bodwin:2016edd}.) In
Refs.~\cite{Bodwin:2007fz,Chung:2010vz}, the scale $\mu$ in
Eq.~(\ref{fVpar-expn}) was taken to be $m_V$. Hence, the values of
$\langle {\cal O}_1^V \rangle_V$ that are given in those papers are
at the scale $m_V$.

The decay rate for the process $H\to V+\gamma$ depends on the decay
constant for a transversely polarized quarkonium $f_V^\perp(\mu)$, which
has the NRQCD expansion \cite{Wang:2013ywc,Bodwin:2014bpa}
\begin{equation}                                                 
f_V^\perp(\mu)                                                 
=
\frac{\sqrt{2N_c}\sqrt{2m_V}}{2m_Q}\Psi_V(0)
\left[1-\frac{5}{6}\langle v^2\rangle_V
+\biggl(-\log\frac{\mu^2}{m_Q^2}-8\biggr)\frac{\alpha_s(\mu) C_F}{4\pi}
+O(\alpha_s^2,\alpha_s v^2,v^4)
\right],
\label{fVperp-expn}
\end{equation}
where $m_Q$ is the heavy-quark pole mass. 

In Ref.~\cite{Bodwin:2016edd}, $f_V^\perp(\mu)$ was assumed to be at the
scale $\mu=m_Q$. However, the values of $\Psi_V(0)$ that were used in
Ref.~\cite{Bodwin:2016edd} were inferred from the values of $\langle
{\cal O}_1^V\rangle$ in Refs.~\cite{Bodwin:2007fz,Chung:2010vz}.
Because those values of $\langle {\cal O}_1^V\rangle$ are at the scale
$\mu=m_V$, it is necessary to correct for the difference in scales.
One can do this straightforwardly by making the following
replacement in the expression for $f_V^\perp$ in 
Eq.~(\ref{fVperp-expn}):
\begin{eqnarray}
\Psi_V(0)\to
\frac{\displaystyle 
1-\frac{1}{6}\langle v^2\rangle_V-8\frac{C_F\alpha_s(m_V)}{4\pi}}
{\displaystyle 
1-\frac{1}{6}\langle v^2\rangle_V-8\frac{C_F\alpha_s(m_Q)}{4\pi}}
\Psi_V(0).
\label{replacement}
\end{eqnarray}

Instead, we follow a procedure that was suggested in
Ref.~\cite{Koenig:2015pha}. We compute $f_V^\parallel$ from the measured
quarkonium electronic decay rates, using Eq.~(\ref{EM-width}). Then, we
use Eq.~(\ref{fVpar-expn}) to eliminate $\Psi_V(0)$ from the amplitude
and use Eqs.~(\ref{fVpar-expn}) and (\ref{fVperp-expn}) to express
$f_V^\perp$ in terms of $f_V^\parallel$:
\begin{subequations}\label{fVpar-fVperp}%
\begin{eqnarray}
f_V^\perp(\mu)&=&\frac{m_V}{2m_Q}
\left[1-\frac{2}{3}\langle v^2\rangle_V
-\frac{\alpha_s(\mu) C_F}{4\pi}\log\frac{\mu^2}{m_Q^2}
+O(\alpha_s^2,\alpha_s v^2,v^4)
\right]f_V^\parallel\label{fVpar-fVperpa} \\
&=&\frac{m_V}{2\overline{m}_Q}
\left[1-\frac{2}{3}\langle v^2\rangle_V
+\frac{\alpha_s(\mu) C_F}{4\pi}\left(
-\log\frac{\mu^2}{\overline{m}_Q^2}-4\right)
+O(\alpha_s^2,\alpha_s v^2,v^4)
\right]f_V^\parallel\label{fVpar-fVperpb}.
\end{eqnarray}
\end{subequations}%
Because the pole mass is subject to renormalon ambiguities and, hence, 
is ill defined, we have used the one-loop relation 
between the $\overline{\rm MS}$ mass $\overline{m}_Q$ 
at the scale $\overline{m}_Q$
and the pole mass
\cite{Tarrach:1980up}, namely,
\begin{equation}
m_Q=\overline{m}_Q\left[1+C_F
\frac{\alpha_s(\overline{m}_Q)}{\pi}+O(\alpha_s^2)\right],
\end{equation}
to express $f_V^\perp/f_V^\parallel$ in terms of the $\overline{\rm MS}$
mass in Eq.~(\ref{fVpar-fVperpb}). As can be seen, if $\mu$ is
approximately equal either to $m_Q$ or to $\overline{m}_Q$, then the
apparent rates of convergence of both the velocity expansions and the
$\alpha_s$ expansions in Eqs.~(\ref{fVpar-fVperpa}) and
(\ref{fVpar-fVperpb}) are better than the corresponding rates of
convergence in Eq.~(\ref{fVperp-expn}). Furthermore, the removal of
$\Psi_V(0)$ from the amplitude reduces some of the theoretical
uncertainties. The resulting expression for the direct amplitude, which
corresponds to Eq.~(37) in Ref.~\cite{Bodwin:2016edd}, is
\begin{eqnarray}
\label{direct-amp}%
i {\cal M}_{\rm dir}^{\rm LC}[\!\!&H&\to V+\gamma]
\nonumber \\
&=& \frac{i}{2} e e_Q \kappa_Q \overline{m}_Q(\mu) 
( \sqrt{2} G_F)^{1/2} 
\left( - \epsilon^*_V \cdot \epsilon^*_\gamma 
+ \frac{\epsilon^*_V \cdot p_\gamma p \cdot \epsilon_\gamma^*}
{p_\gamma \cdot p} \right) 
\frac{f_V^\perp (\mu)}{f_V^\perp (\mu_0)} 
f_V^\parallel \frac{m_V}{2\overline{m}_Q}\nonumber\\
&&\times\left[1-\frac{2}{3}\langle v^2\rangle_V
+\frac{\alpha_s(\mu_0) C_F}{4\pi}
\left(-\log\frac{\mu_0^2}{\overline{m}_Q^2}-4\right)
\right]
\phantom{xxx}
\nonumber \\ && \times
\bigg[ {\cal M}^{(0,0)}(\mu)
+ 
\frac{\alpha_s (\mu)}{4 \pi} {\cal M}^{(1,0)}(\mu)+ 
\frac{\alpha_s (\mu_0)}{4 \pi}{\cal M}^{(0,1)}(\mu)+ 
\langle v^2 \rangle_V {\cal M}^{(0,v^2)}(\mu)
\bigg].
\end{eqnarray}
Here, $e(>0)$ is the electric charge at momentum scale zero, $\kappa_Q$
is a factor that accounts for deviations of the $HQ\bar Q$ coupling from
the standard model (SM) value ($\kappa_Q=1$ is the SM value), $G_F$ is
the Fermi constant, $p$ and $\epsilon^*_V$ are the quarkonium momentum
and polarization, respectively, and $p_\gamma$ and $\epsilon^*_\gamma$
are the photon momentum and polarization, respectively. Each of the
quantities ${\cal M}^{(i,j)}$ is a convolution of a component of the
light-cone hard-scattering kernel at the scale $\mu$ with a component of
the LCDA of the quarkonium, evolved from the scale $\mu_0$ to the
scale $\mu$. The detailed definitions of the ${\cal M}^{(i,j)}$ are
given in Ref.~\cite{Bodwin:2016edd}. As in that work, we carry out the
evolution of the LCDAs at next-to-leading-logarithmic accuracy, using the
Abel-Pad\'e method that is presented in that paper to sum the resulting
series expansions in the eigenfunctions (Gegenbauer polynomials) of the
leading-order evolution operator. Our uncertainties are also computed as
outlined in Ref.\cite{Bodwin:2016edd}.

Note that, because we include the NRQCD expansion for
$f_V^\perp/f_V^\parallel$ as an overall factor in the direct amplitude
in Eq.~(\ref{direct-amp}), this expression contains cross terms of
orders $\alpha_s^2$ and $\alpha_s v^2$ that are beyond the accuracy of
the present calculation.  We structure the amplitude in this way
because of the possibility that the perturbation expansion of
$f_V^\perp/f_V^\parallel$ may be better behaved in higher orders in
$\alpha_s$ than the perturbation expansion of $f_V^\parallel$
\cite{Beneke:1997jm,Czarnecki:2001zc,Beneke:2014qea}.
We use Eq.~(\ref{direct-amp}) to compute the direct amplitude, taking
$\mu=m_H$, as in Ref.~\cite{Bodwin:2016edd}, but taking
$\mu_0=\overline{m}_Q$, rather than $\mu_0=m_Q$ because, as
we have already remarked, the pole mass is ill defined. 

As was pointed out in Ref.~\cite{Bodwin:2013gca}, there is also a
contribution to the decay process from an indirect amplitude, which
arises from the decay of the Higgs boson through a $W$-boson loop or a
quark loop into a photon plus a virtual photon, with the virtual photon
decaying into a quarkonium. We compute the indirect amplitude as in
Ref.~\cite{Bodwin:2016edd}. 

We obtain the results that are shown in Table~\ref{num-amplitudes}
for the direct and indirect amplitudes.
\begin{table}[h]
\begin{center}
\begin{tabular}{lll} 
\hline
$\phantom{xx}V\phantom{x}$ & 
$\phantom{xxxx}\alpha_V$&
$\phantom{xxxxxxxxx}\beta_V$
\\
\hline\hline
$\phantom{x}J/\psi$& 
$11.71\pm 0.17$ & 
$(0.659^{+0.085}_{-0.085}) 
+ (0.073^{+0.035}_{-0.035})i$ \phantom{x}
\\
$\phantom{x}\Upsilon(1S)\phantom{xx}$&
$3.283\pm 0.036\phantom{xxxx}$ &
$(2.962^{+0.078}_{-0.078}) 
+ (0.332^{+0.079}_{-0.079})i$
\\
$\phantom{x}\Upsilon(2S)$ & 
$2.155\pm 0.028$ & 
$(2.072^{+0.056}_{-0.056}) 
+ (0.227^{+0.054}_{-0.054})i$
\\
$\phantom{x}\Upsilon(3S)$ & 
$1.803\pm 0.024$ &
$(1.782^{+0.050}_{-0.050}) 
+ (0.192^{+0.046}_{-0.046})i$
\\
\hline
\end{tabular}
\caption{\label{num-amplitudes}
Values of the parameters $\alpha_V$ and $\beta_V$ in $\Gamma(H\to
V+\gamma)=|\alpha_V-\beta_V\kappa_Q|^{2}\times 10^{-10}~\textrm{GeV}$
for $V=J/\psi$ and $\Upsilon(nS)$. 
}
\end{center}
\end{table}
The results in Table~\ref{num-amplitudes} for the direct amplitude are
shifted by about $5\,\%$ for $H\to J/\psi+\gamma$ and by about $2\,\%$
for $H\to \Upsilon(nS)+\gamma$, relative to the results that were given
in Ref.~\cite{Bodwin:2016edd}. The correction factor in
Eq.~(\ref{replacement}) would shift the direct amplitude by about
$13\,\%$ for $H\to J/\psi+\gamma$ and by about $4$--$5\,\%$ for $H\to
\Upsilon(nS)+\gamma$. Apparently, the additional refinements that we
have introduced in this calculation work to reduce those shifts in the
direct amplitude.

Our results for the decay rates and branching fractions are shown in
Table~\ref{tab:num-rates}.
\begin{table}[h]
\begin{center}
\begin{tabular}{lll} 
\hline
$\phantom{xx}V\phantom{x}$ & 
$\phantom{x}\Gamma(H\to V+\gamma)~(\textrm{GeV})$&
$\phantom{xxx}{\rm Br}(H\to V+\gamma)$
\\
\hline\hline
$\phantom{x}J/\psi$\phantom{x} & 
$\phantom{x}1.221^{+0.042}_{-0.041} \times 10^{-8} $ & 
$\phantom{xxx}2.99^{+0.16}_{-0.15}\times 10^{-6}$ \phantom{x}
\\
$\phantom{x}\Upsilon(1S)\phantom{xx}$&
$\phantom{x}2.13^{+0.82}_{-0.69} \times 10^{-11}\phantom{xx}$ &
$\phantom{xxx}5.22^{+2.02}_{-1.70} \times 10^{-9}$
\\
$\phantom{x}\Upsilon(2S)\phantom{x}$ & 
$\phantom{x}5.82^{+2.93}_{-2.34}\times 10^{-12} $ & 
$\phantom{xxx}1.42^{+0.72}_{-0.57}\times 10^{-9}$
\\
$\phantom{x}\Upsilon(3S)\phantom{x}$ & 
$\phantom{x}3.72^{+1.97}_{-1.55}\times 10^{-12}$ &
$\phantom{xxx}0.91^{+0.48}_{-0.38}\times 10^{-9}$
\\
\hline
\end{tabular}
\caption{\label{tab:num-rates}
SM values of $\Gamma(H\to V+\gamma)$ in units of GeV and ${\rm Br}(H\to V+\gamma)$
for $V=J/\psi$ and $\Upsilon(nS)$. 
}
\end{center}
\end{table}
The results in Table~\ref{tab:num-rates} for the rates are shifted by
about $-1\,\%$ for $H\to J/\psi+\gamma$, by about $-28\,\%$ for $H\to
\Upsilon(1S)+\gamma$, by about $-42\,\%$ for $H\to \Upsilon(2S)+\gamma$,
and by about $-49\,\%$ for $H\to \Upsilon(3S)+\gamma$, relative to the
results that were given in Ref.~\cite{Bodwin:2016edd}. The strong
sensitivity of the rates for $H\to \Upsilon(nS)+\gamma$ to small changes
in the direct amplitude occurs because of the near cancellation of the
direct and indirect amplitudes in these processes.

Given the corrections and refinements to the calculation of the rates
for $H\to V+\gamma$ that we have outlined above, we believe that the
numerical results that we have presented in this paper are more accurate
and reliable than those in Ref.~\cite{Bodwin:2016edd} and should be used
in preference to the results in  Ref.~\cite{Bodwin:2016edd} in future
comparisons with experimental measurements.

\begin{acknowledgments}

The work of G.T.B.\  is supported by the U.S.\ Department of Energy,
Division of High Energy Physics, under Contract No. DE-AC02-06CH11357.
The work of H.S.C.\ at CERN is supported by the Korean Research
Foundation (KRF) through the CERN-Korea fellowship program. The work
of J.-H.E.\ and J.L.\ was supported by 
the National Research Foundation of Korea (NRF) 
under Contract No. NRF-2017R1E1A1A01074699. The submitted manuscript
has been created in part by UChicago Argonne, LLC, Operator of Argonne
National Laboratory. Argonne, a U.S.\ Department of Energy Office of
Science laboratory, is operated under Contract No. DE-AC02-06CH11357.
The U.S.\ Government retains for itself, and others acting on its
behalf, a paid-up nonexclusive, irrevocable worldwide license in said
article to reproduce, prepare derivative works, distribute copies to the
public, and perform publicly and display publicly, by or on behalf of
the Government.

\end{acknowledgments}


\begin{thebibliography}{999}

\bibitem{Bodwin:2016edd} 
  G.~T.~Bodwin, H.~S.~Chung, J.~H.~Ee, and J.~Lee,
\href{http://dx.doi.org/10.1103/PhysRevD.95.054018}
   { Phys.\ Rev.\ D {\bf 95}, no. 5, 054018 (2017)}
  [\href{http://arxiv.org/abs/1603.06793}{arXiv:1603.06793 [hep-ph]}].

\bibitem{Jia:2008ep} 
  Y.~Jia and D.~Yang,
\href{http://dx.doi.org/10.1016/j.nuclphysb.2009.01.025}
     {Nucl.\ Phys.\ B {\bf 814}, 217 (2009)}
  [\href{http://arxiv.org/abs/0812.1965}{arXiv:0812.1965 [hep-ph]}].

\bibitem{Bodwin:1994jh}
  G.\,T.~Bodwin, E.~Braaten, and G.\,P.~Lepage,
\href{http://dx.doi.org/10.1103/PhysRevD.51.1125}
     {Phys.\ Rev.\ D {\bf 51}, 1125 (1995); {\bf 55}, 5853\,(E) (1997)}
  [\href{http://arxiv.org/abs/hep-ph/9407339}{hep-ph/9407339}].

\bibitem{Wang:2013ywc} 
  X.\,P.~Wang and D.~Yang,
\href{http://dx.doi.org/doi:10.1007/JHEP06(2014)121}
     {JHEP {\bf 1406}, 121 (2014)}
  [\href{http://arxiv.org/abs/1401.0122}{arXiv:1401.0122 [hep-ph]}].

\bibitem{Bodwin:2007fz} 
  G.\,T.~Bodwin, H.\,S.~Chung, D.~Kang, J.~Lee, and C.~Yu,
\href{http://dx.doi.org/10.1103/PhysRevD.77.094017}
     {Phys.\ Rev.\ D {\bf 77}, 094017 (2008)}
     [\href{http://arxiv.org/abs/0710.0994}{arXiv:0710.0994 [hep-ph]}].

\bibitem{Chung:2010vz} 
  H.\,S.~Chung, J.~Lee, and C.~Yu,
\href{http://dx.doi.org/10.1016/j.physletb.2011.01.033}
     {Phys.\ Lett.\ B {\bf 697}, 48 (2011)}
  [\href{http://arxiv.org/abs/1011.1554}{arXiv:1011.1554 [hep-ph]}].

\bibitem{Bodwin:2014bpa} 
  G.\,T.~Bodwin, H.\,S.~Chung, J.-H.~Ee, J.~Lee, and F.~Petriello,
\href{http://dx.doi.org/10.1103/PhysRevD.90.113010}
     {Phys.\ Rev.\ D {\bf 90}, 113010 (2014)}
  [\href{http://arxiv.org/abs/1407.6695}{arXiv:1407.6695 [hep-ph]}].

\bibitem{Koenig:2015pha} 
M.~K\"onig and M.~Neubert,
\href{http://dx.doi.org/10.1007/JHEP08(2015)012}
     {JHEP {\bf 1508}, 012 (2015)}
  [\href{http://arxiv.org/abs/1505.03870}{arXiv:1505.03870 [hep-ph]}].

\bibitem{Tarrach:1980up}
  R.~Tarrach,
\href{http://dx.doi.org/10.1016/0550-3213(81)90140-1}
     {Nucl.\ Phys.\ B {\bf 183}, 384 (1981).}
\bibitem{Beneke:1997jm} 
  M.~Beneke, A.~Signer, and V.\,A.~Smirnov,
\href{http://dx.doi.org/10.1103/PhysRevLett.80.2535}
     {Phys.\ Rev.\ Lett.\  {\bf 80}, 2535 (1998)}
  [\href{http://arxiv.org/abs/hep-ph/9712302}{hep-ph/9712302}].
\bibitem{Czarnecki:2001zc} 
  A.~Czarnecki and K.~Melnikov,
\href{http://dx.doi.org/10.1016/S0370-2693(01)01129-7}
     {Phys.\ Lett.\ B {\bf 519}, 212 (2001)}
  [\href{http://arxiv.org/abs/hep-ph/0109054}{hep-ph/0109054}].
\bibitem{Beneke:2014qea} 
  M.~Beneke, Y.~Kiyo, P.~Marquard, A.~Penin, J.~Piclum, D.~Seidel, and 
M.~Steinhauser,
\href{http://dx.doi.org/10.1103/PhysRevLett.112.151801}
     {Phys.\ Rev.\ Lett.\  {\bf 112},  151801 (2014)}
  [\href{http://arxiv.org/abs/1401.3005}{arXiv:1401.3005 [hep-ph]}].

\bibitem{Bodwin:2013gca}
  G.\,T.~Bodwin, F.~Petriello, S.~Stoynev, and M.~Velasco,
\href{http://dx.doi.org/10.1103/PhysRevD.88.053003}{      
  Phys.\ Rev.\ D {\bf 88}, 053003 (2013)}                        
  [\href{http://arxiv.org/abs/1306.5770}{arXiv:1306.5770 [hep-ph]}].

\end{thebibliography}
\end{document}